\documentclass[aps,prl,twocolumn,superscriptaddress,showpacs]{revtex4-1} 
\usepackage{graphicx}
\usepackage{amsmath}
\usepackage{epstopdf}
\usepackage{subfigure}
\usepackage{hyperref}
\begin{document}

\newcommand{\he}{$^4$He } 
\newcommand{\rhos}{$\rho_s$ } 
\newcommand{\<}{\langle}
\renewcommand{\>}{\rangle} 
\newcommand{\Lmin}{L_{\rm min}}
\newcommand{\br}{{\bf r}} 
\newcommand{\bb}{{\bf b}}
\newcommand{\tc}{T_{\text{c}}}
\newcommand{\beq}{\begin{equation}}
\newcommand{\eeq}{\end{equation}}

\title{Superfluid transition in a correlated defect network}

\author{Hannes Meier} 
\author{Mats Wallin} 
\affiliation{Department of Theoretical Physics, KTH-Royal Institute of
Technology, SE-106 91 Stockholm, Sweden}
\author{S.\ Teitel} 
\affiliation{Department of Physics and Astronomy, University of 
Rochester, Rochester, New York 14627, USA} 

\date{\today} 

\begin{abstract} 
Motivated by recent experiments on possible supersolid behavior
  of $^4$He solids at low temperature, we consider a model of
  superfluidity in a defected solid containing a system spaning network of correlated linear dislocations, or planar grain boundaries. 
  Using arguments based on the Harris criterion, as well as numerical simulations,
  we find that such correlated quenched disorder shifts the familiar superfluid lambda transition to a new disordered universality class in which the correlation length exponent $\nu\ge 1$. This results in the temperature-derivates for the superfluid density, $d\rho_{\rm s}/dT$, and for the heat capacity, $dc/dT$, remaining finite at the transition $T_{\rm c}$, and thus a less singular transition, profoundly different from the usual lambda transition.
\end{abstract}

\pacs{64.60.A-, 67.80.bd, 67.25.dj} 

\maketitle

Considerable excitement has been generated by the observation by Kim and Chan \cite{Kim2004a,Kim2004b}, that the period of oscillation of a torsional oscillator (TO) containing solid \he decreases as the system is cooled below roughly 250 mK.  Two different scenarios have been suggested to contribute to this phenomenon.
(i) Initially, it was suggested that a transition to a {\it supersolid} state resulted in a reduction of the mass viscously coupled to the oscillator walls and hence a nonclassical rotational inertia (NCRI).
Subsequent experiments showed this NCRI was closely tied to the presence of defects in the \he crystal:
the slower the crystal was annealed, the smaller was the resulting NCRI signal \cite{Rittner2007}.  
Numerous theoretical calculations and simulations followed, supporting the idea that a supersolid should not exist for a pure \he crystalline state \cite{Prokofevreview,Clark2006,Boninsegni2006}, however superfluidity could exist in the cores of crystalline defects \cite{boninsegni:035301,Soyler2009}.  A system spanning network of pinned defects could then lead to an effective supersolid. 
(ii) Alternatively, it has been suggested that the apparent NCRI is a classical effect due to the viscoelasticity of the defected \he solid \cite{Day,AVB}.  At high temperatures defects are mobile and a fraction of the \he lags the driving oscillation.  But upon cooling, a glassy state is entered in which the defect relaxation time $\tau$ grows dramatically.  When the relaxation rate $1/\tau$ decreases below the TO driving frequency $\omega$, the defects lock into phase with the driving oscillation, leading to a shift in the TO period.

While recent experimental works \cite{Reppy2010,Balibar} seem to favor (ii) as the dominant mechanism for the observed NCRI, 
this scenario still assumes a highly defected crystal; one might therefore still expect to see some contribution from superfluidity in the defect network, as in scenario (i).  
While direct probes of a steady state superfluid flow have yielded conflicting results \cite{Ray:2008p638}, other 
experiments have failed to see the  characteristic features expected for the bulk lambda transition of three dimensional ($d=3$) superfluid  $^4$He \cite{spaceShuttle}.  In particular, (1) the TO period shift varies rather smoothly with temperature  \cite{Kim2004a,Kim2004b,West,ClarkWestChan2007}, with no sign of the sharp singularity expected if some part of this shift was due to a superfluid density $\rho_{\rm s}\sim |T-T_{\rm c}|^\nu$, with $\nu\approx 0.67$; and (2) equilibrium measurements of heat capacity  \cite{Lin2007,Lin2009} show a smooth bump rather than the sharp lambda-shaped cusp characteristic of a bulk superfluid transition with $c\sim|T-T_{\rm c}|^{-\alpha}$, $\alpha=2-d\nu\approx -0.01$ \cite{Hasenbusch}.  One possible reason for the absence of such sharp features is if the dislocation density is simply too small to give an observable superfluid signal \cite{boninsegni:035301}.  In this work, however, we suggest another possible contributing factor: that the form of the disorder may shift the superfluid transition to a new universality class, with  much less singular behavior.

The Harris criterion \cite{Harris} argues that uncorrelated point-like disorder is irrelevant whenever $2<d\nu$, with $d$ the spatial dimension of the system, and $\nu$ the correlation length critical exponent.  For bulk superfluidity in $d=3$, the pure system has $\nu_{\rm pure}\approx 0.67$ and the Harris criterion is satisfied.  It has therefore been argued \cite{Toner2008,Goswami2011, ProkofevRMP2012} that  the superfluid onset in a random dislocation network should show the same lambda singularity as in bulk superfluid helium.
However, since the  dislocation cores are continuous one dimensional objects, the assumption of uncorrelated {\it point-like} disorder might not be appropriate.  
In this work we consider the case when disorder is correlated along a network of intersecting lines or planes.
In all cases we argue that the effects of such correlated  disorder is to lessen the sharpness of the critical singularity, removing the divergent temperature-derivatives at $T_{\rm c}$ that are characteristic of the usual bulk lambda transition.

To model superfluidity in the dislocation cores of a defected \he crystal, we start with a lattice 3D XY model. Neglecting amplitude fluctuations of the condensate order parameter $\Psi(\br)=|\Psi|e^{i\theta(\br)}$,
the fluctuations in the phase angle $\theta$ that drive the superfluid transition can be
modeled by the Hamiltonian,
\beq
{\cal H}=-\sum_{i,\mu} J_\mu(\br_i) \cos[\theta(\br_i+\hat\mu)-\theta(\br_i)] \enspace,
\label{H}
\eeq
where $i$ labels the discrete sites ${\br_i}$ of a simple cubic lattice of length $L$ with periodic boundary
conditions in all directions, and $J_\mu(\br_i)$ is the coupling on the bond connecting site $\br_i$ to $\br_i+\hat\mu$, with
$\mu=x, y, z$ the lattice axis directions.  We will choose spatially inhomogeneous random couplings $J_\mu(\br_i)$ according to several different schemes.  Although in  real solid \he dislocation cores can thermally fluctuate at higher temperatures, we will assume that at sufficiently low $T$ the dislocations are pinned by $^3$He impurities, and so we take the couplings $J_\mu(\br_i)$ to be fixed quenched variables.

According to Harris \cite{Harris}, the effect of disorder is controlled by 
the mean square fluctuation of the local coupling $\Delta J$, disorder averaged over a sub-volume $V=R^d$ of the system,
\beq
\label{DJ}
[(\Delta J)^2_R]\equiv \left[\left\{R^{-d}\sum_\mu\sum_{\br_i\in V} (J_\mu(\br_i)-[J_\mu])\right\}^2\right]\enspace,
\eeq
where $[\dots ]$ denotes the average over different realizations of quenched disorder.  For a continuous transition with diverging correlation length $\xi\sim|T-T_{\rm c}|^{-\nu}$, the Harris criterion argues that disorder is irrelevant at  $T_{\rm c}$ when the coupling fluctuation averaged over a correlation volume $\xi^d$ satisfies $\sqrt{[(\Delta J)^2_\xi]} <|T-T_{\rm c}|$ as $T\to T_{\rm c}$.  For future use, we note that $[(\Delta J)^2_R]$ can be rewritten in terms of the  coupling correlation averaged over the sub-volume $V$,
\beq
[(\Delta J)^2_R]=C(R)\equiv R^{-d}\sum_{\mu,\mu^\prime}\sum_{\br_i\in V}[\delta J_\mu(\br_i)\delta J_{\mu^\prime}(0)]\enspace,
\label{DJ2}
\eeq
where $\delta J_\mu(\br_i)\equiv J_\mu(\br_i)-[J_\mu]$.  
For uncorrelated point disorder, with $[\delta J_\mu(\br_i)\delta J_{\mu^\prime}(0)]\propto \delta_{\mu,\mu^\prime}\delta_{\br_i,0}$, Eq.~(\ref{DJ2}) yields $C(\xi)\sim\xi^{-d}\sim |T-T_{\rm c}|^{d\nu}$.  The Harris criterion for the irrelevance of such point disorder then becomes $|T-T_{\rm c}|^{d\nu/2}<|T-T_{\rm c}|$ or the familiar $2<d\nu$.

We now wish to consider models of correlated disorder.  One such case was considered many years ago in a seminal work by Weinrib and Halperin \cite{Weinrib}.  They noted that when spatial disorder is introduced as a set of straight {\it randomly oriented} lines, the resulting disorder averaged coupling correlation decays algebraically as $[\delta J({\bf r})\delta J(0)]\sim r^{-a}$, with $a=d-1$.  In this case, Eq.~(\ref{DJ2}) gives $C(\xi)\sim \xi^{-a}$ and the Harris criterion argues that disorder is irrelevant when $2<a\nu$.
Using a renormalization group expansion for weak Gaussian disorder, Weinrib and Halperin showed that this is indeed the case:  when $a<d$, disorder is irrelevant whenever $2<a\nu_{\rm pure}$.  They further found that when this criterion fails, the correlation length exponent at the new disordered critical point satisfies $\nu_{\rm disor}=2/a$.  
Applying this conclusion to a 3D superfluid, we see that such linear disorder with $a=2$ is relevant, resulting in a superfluid density that vanishes linearly as $\rho_{\rm s}\sim |T-T_{\rm c}|^{\nu_{\rm disor}}$, with $\nu_{\rm disor}=1$, and a heat capacity  $c\sim|T-T_{\rm c}|^{-\alpha}$ with exponent $\alpha=2-d\nu_{\rm disor}=-1$.  In particular,  $dc/dT$ scales as $|T-T_{\rm c}|^{-\alpha-1}$, with $-\alpha-1=0$, and thus remains finite at $T_{\rm c}$ in contrast to its divergence in the pure model.  Thus for such linearly correlated disorder,  we expect that the superfluid density vanishes  less sharply than the pure model, while the heat capacity has a far less singular cusp.  For planar disorder, as might be the case if superfluidity is carried on randomly oriented twin grain boundaries, $a=d-2=1$ in 3D, and the superfluid singularities are even less sharp.

We next consider a slightly different model of  linearly correlated disorder.  Motivated by the notion that dislocation lines in solid $^4$He may preferentially align parallel to crystalline axes, we consider a 3D XY model in which correlated disorder enters along straight lines oriented only along the three lattice directions.
We choose quenched couplings as follows.  Within the plane at $r_\mu=0$, we randomly choose equal numbers of bonds $J_\mu$  from the bi-valued distribution $J=1\pm\delta$.  We then continue these couplings in correlated straight lines by requiring $J_\mu(\br_i+\hat\mu)=J_\mu(\br_i)$.  We follow this procedure for all three directions $\mu=x,y,z$. 
For the results presented below we use $\delta=1$; superfluidity is strictly confined to the dislocation cores and the three dimensionality of the superflow results solely from the intersections of these cores to form an interconnected network.  We have also considered the case $\delta=0.95$, as a model in which superfluid particles may tunnel through the bulk between dislocation cores.  We find the critical behavior to be the same in both cases.  The coupling correlation for this disorder is $[\delta J_\mu(\br)\delta J_{\mu^\prime}(0)]\propto[\delta(x)\delta(y)+\delta(y)\delta(z)+\delta(z)\delta(x)]\delta_{\mu,\mu^\prime}$. However the volume averaged correlation is $C(R)\sim R^{-a}$ with $a=d-1$, just as for randomly oriented lines. We thus expect this disorder to be relevant; applying the Harris criterion at the new disordered critical point then requires $\nu_{\rm disor}\ge 2/a=1$. 

To confirm this behavior, we carry out extensive numerical simulations.  The superfluid density is proportional to the XY helicity modulus,
which (in units where $m/\hbar=1$) is given by \cite{Li:1990p4584}
\beq 
\label{rho_s}
\begin{split}
\rho_{\rm s} = L^{-3} \left[ \left\<  \sum_i J_z(\br_i) \cos[\theta(\br_i+\hat z)-\theta(\br_i)] \right\> \right. \\
- T^{-1} \left. \left\< \left( \sum_i J_{z}(\br_i) \sin[\theta(\br_i+\hat z)-\theta(\br_i)] \right)^2 \right\> \right]\enspace.
\end{split} 
\eeq
The XY magnetization is  $M=  \left| \sum_i  e^{i\theta(\br_i)} \right|$, and its
Binder cumulant $U$ is given by,
\beq 
\label{U}
U\equiv\left[ \frac{\<M^4\>}{\<M^2\>^2} \right]\enspace.
\eeq 
We also consider the heat capacity $c$, 
\beq
\label{c}
c=L^{-3} T^{-2} \left[ \<{\cal H}^2\>  - \<{\cal H}\>^2 \right] \enspace,
\eeq
and the XY spin susceptibility $\chi$, 
\beq
\label{chi}
\chi = L^{-3} T^{-1} \left[ \<M^2\>  - \<M\>^2 \right] \enspace.
\eeq

To analyze the critically of the transition at $T_{\rm c}$ we apply finite-size-scaling (FSS) methods \cite{FSS}.  We expect $\rho_{\rm s}$, $U$, and $\chi$ as a function of $T$ and system length $L$,  to obey the usual FSS relations,
\beq
\begin{split}
\rho_{\rm s}(T,L) =L^{-1}{\cal R}(tL^{1/\nu})\enspace,
\quad
U(T,L)={\cal U}(tL^{1/\nu})\enspace,
\\
\chi(T,L)= L^{\gamma/\nu} {\cal X}(tL^{1/\nu})\enspace,\qquad\qquad\quad
\end{split}
\label{all_scal}
\eeq
where ${\cal R}(\cdot)$, ${\cal U}(\cdot)$ and ${\cal X}(\cdot)$ are scaling functions, and $t\equiv(T-T_{\rm c})/T_{\rm c}$.

To carry out our Monte Carlo (MC) simulations to high accuracy we used an effective
Wolff collective update algorithm, which minimizes the effects of critical
slowing down at the transition \cite{wolff}.  As the Wolff algorithm is not so efficient at lower temperatures, we add one ordinary Metropolis sweep through the system for each 100 Wolff updates.  Data for the averages
were accumulated during $4-16\times 10^{3}$ Monte Carlo steps (MCS), after
equally many initial MCS were discarded to reach equilibrium.  One MCS
is defined as $L^3$ single site update attempts.  We tested for
equilibration by increasing the number of discarded initial MCS until
stable results were obtained.  For our FSS analysis, quenched disorder averages were computed
over $10^3 - 10^{4}$ independent realizations of the random couplings
$J_\mu(\br_i)$.

\begin{figure}[h!]
\includegraphics[width=0.48\textwidth]{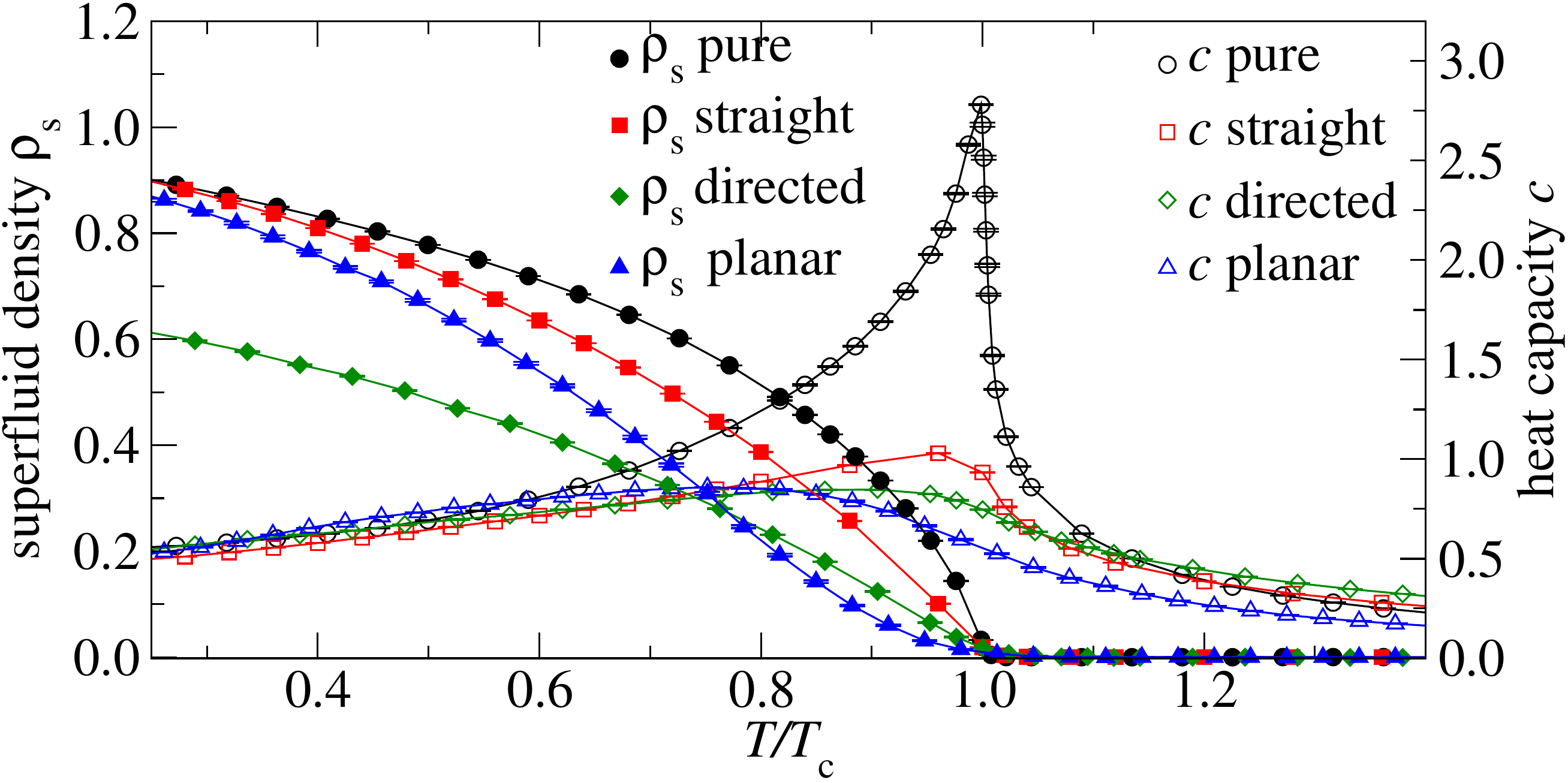}
\caption[]{(color online) Superfluid density $\rho_{\rm s}$ (solid symbols) and heat capacity $c$ (open symbols), vs $T/T_{\rm c}$, for several model systems on a lattice of length $L=40$. Black circles: pure 3D XY model; red squares: straight line disorder; green diamonds: directed random line disorder; blue triangles: random planar disorder.
For the disordered models, results are averaged over several hundred to more than a thousand random coupling configurations}
\label{fig1} 
\end{figure}

The main qualitative results of this work are shown in Fig.\
\ref{fig1}, where we plot the superfluid density $\rho_{\rm s}$ and heat capacity $c$ vs $T/T_{\rm c}$ of a system of fixed length $L=40$, for both
the pure 3D XY model ($T_{\rm c}=2.203$) and for our model of Eq.~(\ref{H}) with linearly correlated disorder ($T_{\rm c}=2.501$, as determined below).  Results for the disordered case are averaged over several hundred realizations of the quenched random couplings.
We see that the presence of the disorder dramatically softens the singularities of the pure system, removing the sharp singularities of the lambda transition.  Derivatives with respect to temperature that diverge as $T\to T_{\rm c}$ in the pure model, appear to become finite in the disordered model.

\begin{figure} [h!]
\includegraphics[width=0.48\textwidth]{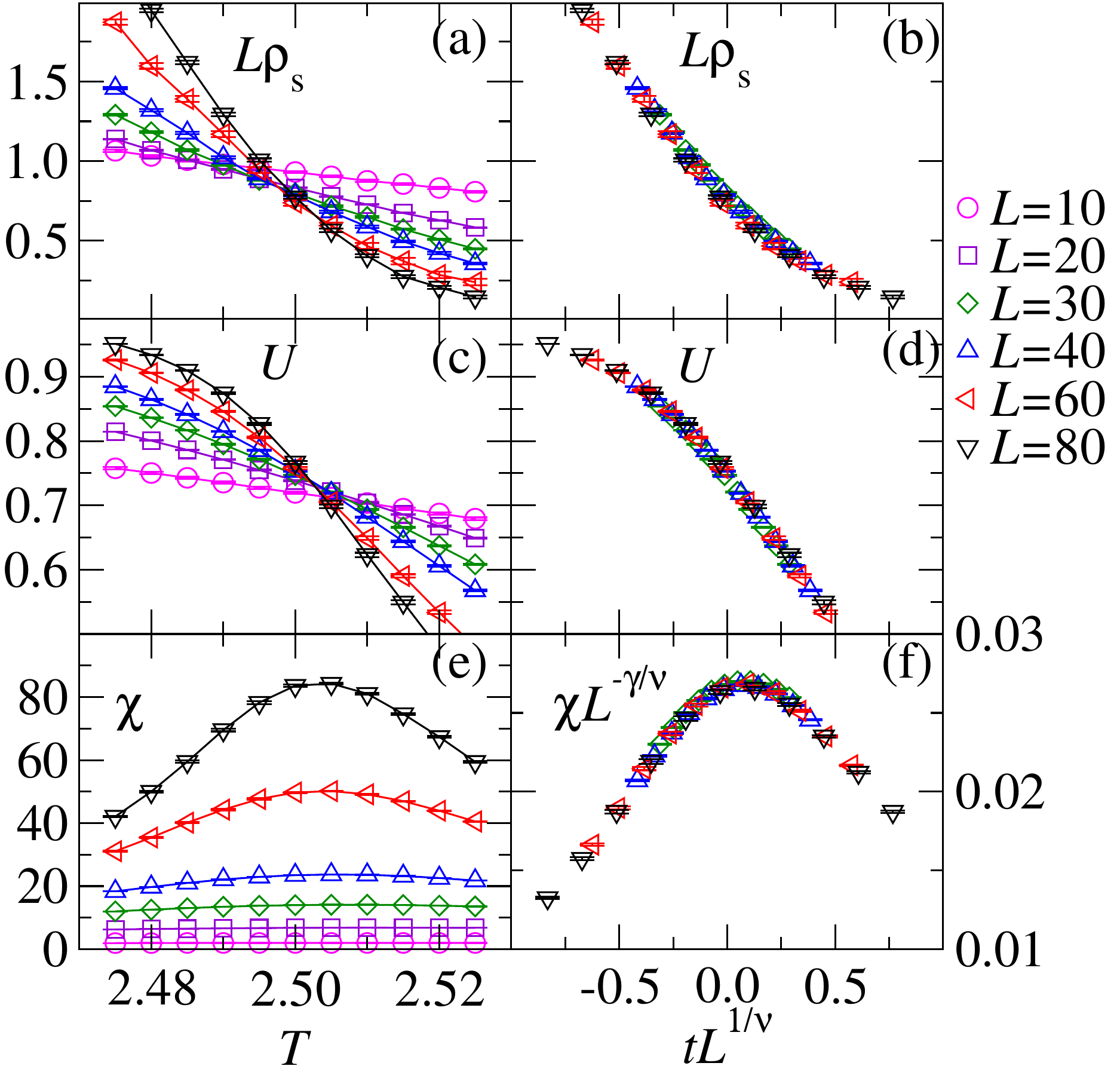}
\caption[]{(color online) (a) Scaled superfluid density $L\rho_{\rm s}$, (c) Binder cumulant $U$, and (e) susceptibility $\chi$ vs $T$ for system sizes $L=10-80$.  Scaling collapse of (b) $L\rho_{\rm s}$, (d) $U$, and (f) $\chi L^{-\gamma/\nu}$ vs $tL^{1/\nu}$, for system sizes $L=30-80$, using values $T_{\rm c}=2.501$, $\nu=1$ and $\gamma=1.84$. 
}
\label{fig2} 
\end{figure}

To verify this conclusion, and to demonstrate that the model with linearly correlated defects belongs to a new disordered universality class, we perform a FSS analysis to determine critical exponents
\cite{FSS}. In Figs.\ \ref{fig2}(a) and (c) we plot respectively $L\rho_{\rm s}$ and $U$ vs $T$ for different system sizes $L=10-80$.  Equations\ (\ref{all_scal}) predict that the curves for different $L$ should all intersect at the common point $t=0$, i.e. when $T=T_{\rm c}$.
We see that as $L$ increases, the curves do indeed seem to be approaching a common intersection point, yielding $T_{\rm c}\approx 2.5$.   The deviations from a perfect common intersection are due to corrections to scaling, which can be noticeable when $L$ is insufficiently large \cite{Hasenbusch}.  

\begin{figure} [h!]
\includegraphics[width=0.48\textwidth]{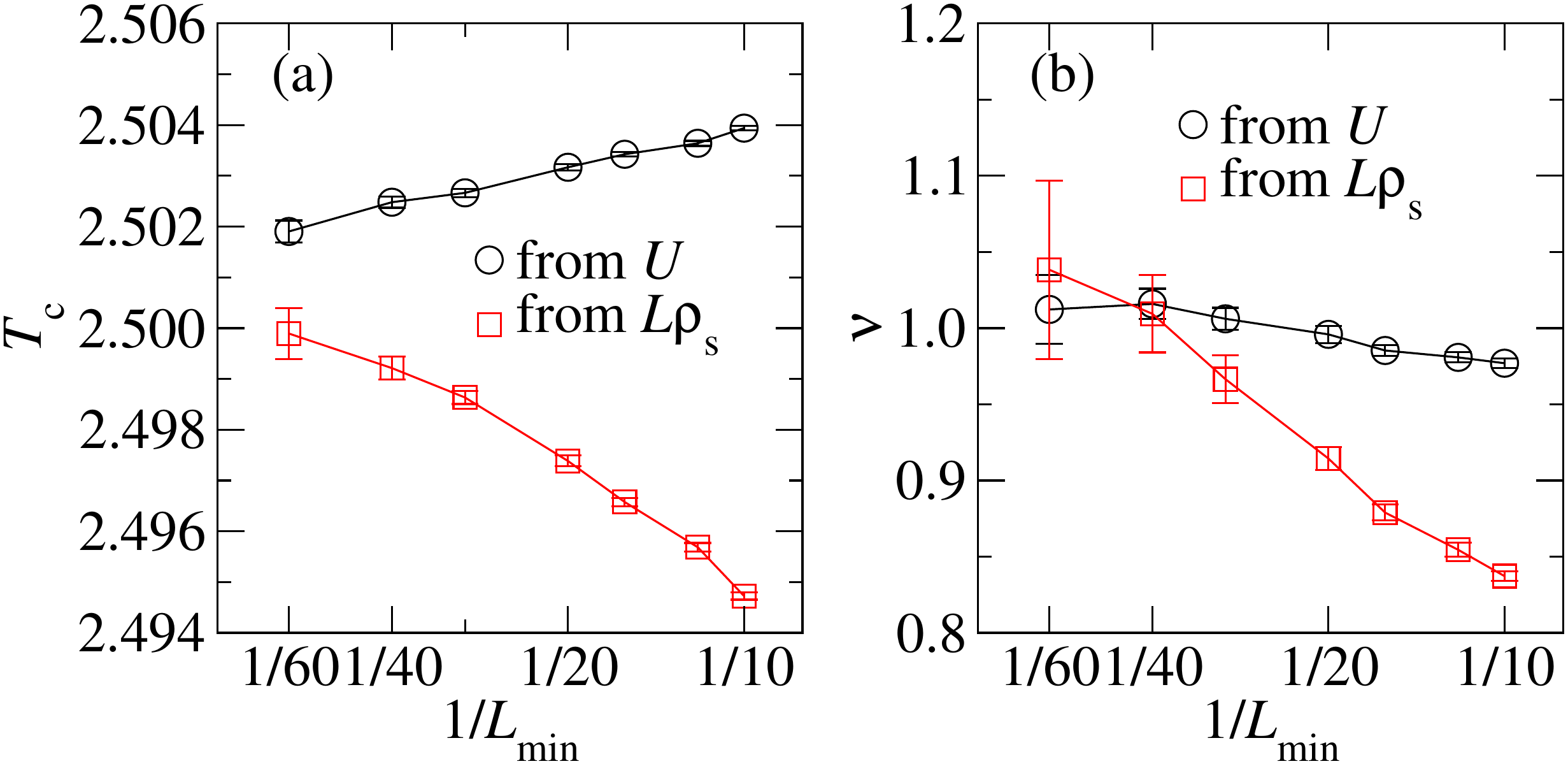}
\caption[] {(color online) Values of (a) $T_{\rm c}$ and (b) $\nu$ obtained by fitting data with $|tL^{1/\nu}|<0.5$ to third order polynomial expansions of the scaling functions of Eqs.\ (\ref{all_scal}), using data from system sizes $L_{\rm min}$ to $L_{\rm max}=80$.
}
\label{fig3}
\end{figure}

For a more accurate determination of the critical $T_{\rm c}$, as well as the correlation length exponent $\nu$, we expand the scaling functions of Eqs.\ (\ref{all_scal}) as third order polynomials, and fit our data to these scaling forms with $T_{\rm c}$, $\nu$ and the polynomial coefficients as free fitting parameters.  We restrict our fits to data satisfying $|tL^{1/\nu}|<0.5$.  Using system sizes $L_{\rm min}$ to $L_{\rm max}=80$, we plot the resulting fitted values for $T_{\rm c}$ and $\nu$ vs $1/L_{\rm min}$ in Figs.~\ref{fig3}(a) and (b) respectively.  We see that as $L_{\rm min}$ increases, the values of $T_{\rm c}$ and $\nu$ from $L\rho_{\rm s}$ and $U$ approach each other.  A procedure {\it including} corrections to scaling yields consistent values of $T_{\rm c}=2.501\pm 0.001$ and $\nu=1.00\pm 0.05$ from fits to both $L\rho_{\rm s}$ and $U$.  Errors represent one standard deviation statistical error as estimated using the method of synthetic data sets \cite{NumericalRecipesChpt15.6}.  
Using these common values of $T_{\rm c}$ and $\nu$, the resulting scaling collapses, plotting $L\rho_{\rm s}$ and $U$ vs $tL^{1/\nu}$, are shown in Figs.\ \ref{fig2}(b) and (d), for sizes $L=30-80$.  
We note that our result $\nu \approx 1$ is identical to the prediction of Weinrib and Halperin for algebraically correlated couplings with $a=d-1=2$, and implies that the new disordered fixed point just marginally satisfies the Harris criterion for stability.

Next we consider the susceptibility $\chi$ of Eq.\ (\ref{chi}).  In Fig.\ \ref{fig2}(e) we plot $\chi$ vs $T$ for system sizes $L=10-80$.  The exponent
$\gamma$ in Eq.\ (\ref{all_scal}) is obtained from a power law fit to the
maximum value of $\chi(T)$ vs $L$.  We find
$\gamma=1.84 \pm 0.1$. 
Using this value of $\gamma$, and $T_{\rm c}=2.501, \nu=1.0$ as determined above, 
we show in Fig.\ \ref{fig2}(f) the resulting scaling collapse, plotting $\chi L^{-\gamma/\nu}$ vs $tL^{1/\nu}$ for system sizes $L=30-80$.  The collapse is excellent.
From the two independent exponents $\nu \approx
1.0, \gamma \approx 1.84$ all the other thermodynamic critical
exponents can be obtained from standard scaling laws \cite{Ma}.  Hyperscaling gives the heat capacity exponent
$\alpha=2-d\nu\approx -1.0$; the Rushbrook equality, $\alpha+2\beta+\gamma=2$, gives for the order parameter exponent $\beta\approx 0.58$; and the correlation function exponent $\eta$ is obtained from
$\gamma=(2-\eta)\nu$ giving $\eta \approx 0.16$.

\begin{figure} [h!]
\includegraphics[width=0.48\textwidth]{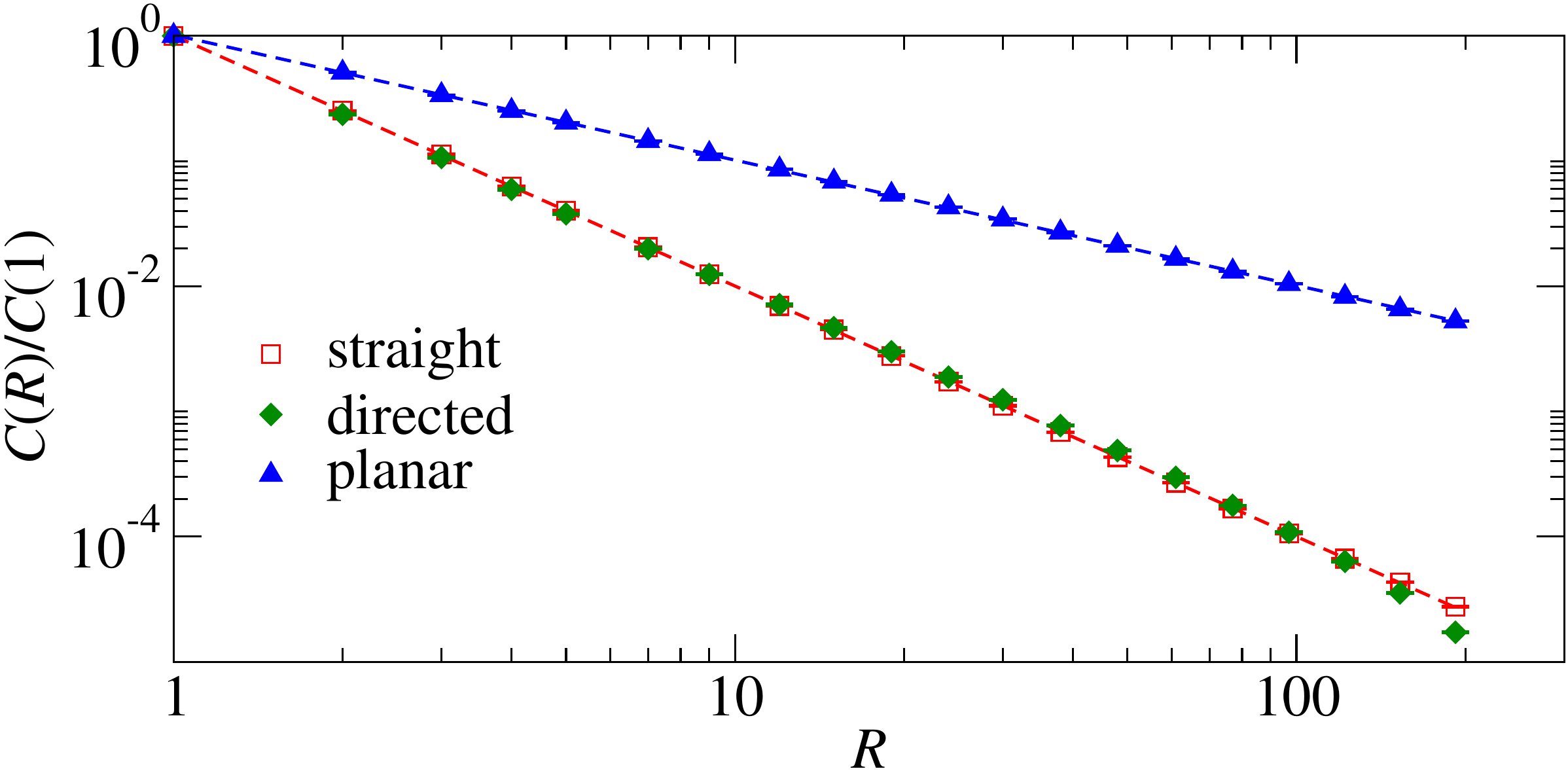}
\caption[] {(color online) Integrated coupling correlation $C(R)$ of Eq.~(\protect\ref{DJ2}), vs $R$, for a system of length $L=200$.  Results are normalized by $C(1)$, and shown for the case of straight randomly positioned lines, directed random lines, and random flat planes.
}
\label{fig4}
\end{figure}

So far, our linear disorder has taken the form of perfectly straight lines.  In a solid, however, dislocation cores may wander as they traverse the system.  To consider the effect of such wandering, we generalize our model by letting each defect line be a directed random walk. For a walk directed along $\hat z$, for example, each step in the $\hat z$ direction is allowed to include a random transverse fluctuation $\Delta {\bf r}_\perp$, sampled equally from $\Delta \br_\perp=\{0,\pm\hat x,\pm\hat y\}$.  For each walk the $\Delta {\bf r}_\perp$ are constrained to sum to zero, so that the line at $z=L$ returns to its starting position at $z=0$ under the periodic boundary conditions.  
For each such walk $j$, a variable $n_j=1$ is placed on each bond of the walk.
The couplings of the corresponding XY model are then set to $J=\sum_j n_j$ on each bond.  
Finally, all couplings $J$ are rescaled by a constant factor so that the disorder average $[J_\mu]=1$.
We thus construct such configurations on a lattice of size $L=200$ and numerically compute the integrated coupling correlation $C(R)$ of Eq.~(\ref{DJ2}), averaging over more than 1000 different realizations of the disorder.  Our results are shown in Fig.~\ref{fig4}, along with $C(R)$ as computed for our original straight line model.  We find a clear algebraic decay in both cases, with $C(R)\sim R^{-a}$, $a=2$.  Thus the directed random line model is expected to be in the same universality class as our straight line model.  In Fig.~\ref{fig1} we show results for $\rho_{\rm s}$ and heat capacity $c$ for this directed random line model, as obtained from simulations on a lattice of size $L=40$, averaging over several hundreds of disorder configurations.  The effect of superfluidity seems slightly suppressed, as compared to the straight line case, but the shapes of the singularities appear to be the same.

We have also considered a random plane model.  We select a sequence of flat planes as follows.  For planes oriented with normal along $\hat z$, each $xy$ plane at height $z$ is randomly selected or not with probability $1/2$.  Such random plane sequences are selected in all three directions $\hat x$, $\hat y$, $\hat z$ simultaneously. 
For each such plane $j$ we set $n_j=1$ for all bonds in the plane; the XY bond couplings are $J=\sum_j n_j$, and then rescaled so that $[J]=1$.
In Fig.~\ref{fig4} we plot the resulting numerically computed correlation $C(R)$.  We find an algebraic decay with $C(R)\sim R^{-a}$, $a=1$. In Fig.~\ref{fig1} we show results for $\rho_{\rm s}$ and heat capacity $c$ for this random plane model, as obtained from simulations on a lattice of size $L=40$, averaging over more than one thousand disorder configurations.  As expected, the singularity at $T_{\rm c}$ is now even smoother than for the random line models.

To conclude, we have considered a variety of models for a superfluid with long range correlated quenched disorder.  In all cases we have argued that the disorder changes the universality class of the superfluid transition to one with $\nu\ge 1$.  Our results show that superfluidity in such correlated disorder networks, such as have been proposed in some models for solid $^4$He, would not display the familiar sharp features of the lambda transition of an ordinary bulk superfluid.

We are grateful to Egor Babaev, Alexander Balatsky, and John Reppy for
discussions. This work was supported by the Swedish Research Council, the G{\"o}ran Gustafsson foundation, NSF grant DMR-1205800
and the Swedish National Infrastructure for Computing (SNIC).


\bigskip





\begin{thebibliography}{10}%


\makeatletter
\providecommand \@ifxundefined [1]{%
 \ifx #1\undefined \expandafter \@firstoftwo
 \else \expandafter \@secondoftwo
\fi
}%
\providecommand \@ifnum [1]{%
 \ifnum #1\expandafter \@firstoftwo
 \else \expandafter \@secondoftwo
\fi
}%
\providecommand \enquote [1]{``#1''}%
\providecommand \bibnamefont  [1]{#1}%
\providecommand \bibfnamefont [1]{#1}%
\providecommand \citenamefont [1]{#1}%
\providecommand\href[0]{\@sanitize\@href}%
\providecommand\@href[1]{\endgroup\@@startlink{#1}\endgroup\@@href}%
\providecommand\@@href[1]{#1\@@endlink}%
\providecommand \@sanitize [0]{\begingroup\catcode`\&12\catcode`\#12\relax}%
\@ifxundefined \pdfoutput {\@firstoftwo}{%
 \@ifnum{\z@=\pdfoutput}{\@firstoftwo}{\@secondoftwo}%
}{%
 \providecommand\@@startlink[1]{\leavevmode}%
 \providecommand\@@endlink[0]{}%
}{%
 \providecommand\@@startlink[1]{%
  \leavevmode
  \pdfstartlink
   attr{/Border[0 0 1 ]/H/I/C[0 1 1]}%
   user{/Subtype/Link/A<</Type/Action/S/URI/URI(#1)>>}%
  \relax
 }%
 \providecommand\@@endlink[0]{\pdfendlink}%
}%
\providecommand \url  [0]{\begingroup\@sanitize \@url }%
\providecommand \@url [1]{\endgroup\@href {#1}{\urlprefix}}%
\providecommand \urlprefix [0]{URL }%
\providecommand \Eprint[0]{\href }%
\@ifxundefined \urlstyle {%
  \providecommand \doi [1]{doi:\discretionary{}{}{}#1}%
}{%
  \providecommand \doi [0]{doi:\discretionary{}{}{}\begingroup
  \urlstyle{rm}\Url }%
}%
\providecommand \doibase [0]{http://dx.doi.org/}%
\providecommand \Doi[1]{\href{\doibase#1}}%
\providecommand \bibAnnote [3]{%
  \BibitemShut{#1}%
  \begin{quotation}\noindent
    \textsc{Key:}\ #2\\\textsc{Annotation:}\ #3%
  \end{quotation}%
}%
\providecommand \bibAnnoteFile [2]{%
  \IfFileExists{#2}{\bibAnnote {#1} {#2} {\input{#2}}}{}%
}%
\providecommand \typeout [0]{\immediate \write \m@ne }%
\providecommand \selectlanguage [0]{\@gobble}%
\providecommand \bibinfo [0]{\@secondoftwo}%
\providecommand \bibfield [0]{\@secondoftwo}%
\providecommand \translation [1]{[#1]}%
\providecommand \BibitemOpen[0]{}%
\providecommand \bibitemStop [0]{}%
\providecommand \bibitemNoStop [0]{.\EOS\space}%
\providecommand \EOS [0]{\spacefactor3000\relax}%
\providecommand \BibitemShut [1]{\csname bibitem#1\endcsname}%




\bibitem{Kim2004a}%
  \BibitemOpen
  \bibfield{author}{%
  \bibinfo {author} {\bibfnamefont{E.}~\bibnamefont{Kim}}\ and\ \bibinfo
  {author} {\bibfnamefont{M.~H.~W.}\ \bibnamefont{Chan}},\ }%
  \bibfield{journal}{%
  \bibinfo {journal} {Science}\ }%
  \textbf{\bibinfo {volume} {305}},\ 
\bibinfo {pages} {1941}  
(\bibinfo {year} {2004}).





\bibitem{Kim2004b}%
  \BibitemOpen
  \bibfield{author}{%
  \bibinfo {author} {\bibfnamefont{E.}~\bibnamefont{Kim}}\ and\ \bibinfo
  {author} {\bibfnamefont{M.~H.~W.}\ \bibnamefont{Chan}},\ }%
  \bibfield{journal}{%
  \bibinfo {journal} {Nature}\ }%
  \textbf{\bibinfo {volume} {427}},\ 
\bibinfo {pages} {225}  
(\bibinfo {year} {2004}).


\bibitem{Rittner2007}%
  \BibitemOpen
  \bibfield{author}{%
  \bibinfo {author} {\bibfnamefont{A.~S.~C.}\ \bibnamefont{Rittner}}\ and\
  \bibinfo {author} {\bibfnamefont{J.~D.}\ \bibnamefont{Reppy}},\ }%
  \bibfield{journal}{%
  \bibinfo {journal} {\prl}\ }%
  \textbf{\bibinfo {volume} {98}},\ 
\bibinfo {pages} {175302} 
   (\bibinfo {year} {2007}).


\bibitem{Prokofevreview}%
  \BibitemOpen
  \bibfield{author}{%
  \bibinfo {author} {\bibfnamefont{N.~V.}~\bibnamefont{Prokof'ev}},\ }%
  \bibfield{journal}{%
  \bibinfo {journal} {Advances in Physics}\ }%
  \textbf{\bibinfo {volume} {56}},\ 
\bibinfo {pages} {381} 
   (\bibinfo {year} {2007}).



\bibitem{Clark2006}%
  \BibitemOpen
  \bibfield{author}{%
  \bibinfo {author} {\bibfnamefont{B.~K.}~\bibnamefont{Clark}}\ and\ \bibinfo
  {author} {\bibfnamefont{D.~M.}~\bibnamefont{Ceperley}},\ }%
  \bibfield{journal}{%
  \bibinfo {journal} {\prl}\ }%
  \textbf{\bibinfo {volume} {96}},\ 
\bibinfo {pages} {105302} 
(\bibinfo {year} {2006}).



\bibitem{Boninsegni2006}%
  \BibitemOpen
  \bibfield{author}{%
  \bibinfo {author} {\bibfnamefont{M.}~\bibnamefont{Boninsegni}}, \bibinfo
  {author} {\bibfnamefont{N.~V.}~\bibnamefont{Prokof'ev}},\ and\ \bibinfo {author}
  {\bibfnamefont{B.~V.}~\bibnamefont{Svistunov}},\ }%
  \bibfield{journal}{%
  \bibinfo {journal} {\prl}\ }%
  \textbf{\bibinfo {volume} {96}},\ 
\bibinfo {pages} {105301} 
(\bibinfo {year} {2006}).


\bibitem{boninsegni:035301}%
  \BibitemOpen
  \bibfield{author}{%
  \bibinfo {author} {\bibfnamefont{M.}~\bibnamefont{Boninsegni}}, \bibinfo
  {author} {\bibfnamefont{A.~B.}\ \bibnamefont{Kuklov}}, \bibinfo {author}
  {\bibfnamefont{L.}~\bibnamefont{Pollet}}, \bibinfo {author}
  {\bibfnamefont{N.~V.}\ \bibnamefont{Prokof'ev}}, \bibinfo {author}
  {\bibfnamefont{B.~V.}\ \bibnamefont{Svistunov}},\ and\ \bibinfo {author}
  {\bibfnamefont{M.}~\bibnamefont{Troyer}},\ }%
  \bibfield{journal}{%
  \bibinfo {journal} {\prl}\ }%
  \textbf{\bibinfo {volume} {99}},\ 
\bibinfo {pages} {035301} 
(\bibinfo {year} {2007}).



\bibitem{Soyler2009}
S.~G.\ S\"oyler, A.~B.\ Kuklov, L.\ Pollet, N.~V.\ Prokof'ev, and B.~V.\ Svistunov,
\prl {\bf 103}, 175301 (2009).


\bibitem{Day}%
  \BibitemOpen
  \bibfield{author}{%
  \bibinfo {author} {\bibfnamefont{J.}~\bibnamefont{Day}}\ and\ \bibinfo
  {author} {\bibfnamefont{J.}~\bibnamefont{Beamish}},\ }%
  \bibfield{journal}{%
  \bibinfo {journal} {Nature}\ }%
  \textbf{\bibinfo {volume} {450}},\ 
\bibinfo {pages} {853} 
(\bibinfo {year} {2007}).



\bibitem{AVB}
A.~V.\ Balatsky, M.~J.\ Graf, Z.\ Nussinov, and J.-J.\ Su, arXiv:1209.0803.




\bibitem{Reppy2010}%
  \BibitemOpen
  \bibfield{author}{%
  \bibinfo {author} {\bibfnamefont{J.~D.}~\bibnamefont{Reppy}},\ }%
  \bibfield{journal}{%
  \bibinfo {journal} {\prl}\ }%
  \textbf{\bibinfo {volume} {104}},\ 
\bibinfo {pages} {255301} 
(\bibinfo {year} {2010}).


\bibitem{Balibar}A.~D.\ Fefferman, X.\ Rojas, A.\ Haziot, S.~E.\ Balibar, 
J.~T.\ West, and M.~H.~W.\ Chan, \prb {\bf 85}, 094103 (2012).


\bibitem{Ray:2008p638}%
  \BibitemOpen
  \bibfield{author}{%
  \bibinfo {author} {\bibfnamefont{M.~W.}~\bibnamefont{Ray}}\ and\ \bibinfo
  {author} {\bibfnamefont{R.~B.}~\bibnamefont{Hallock}},\ }%
  \bibfield{journal}{%
  \bibinfo {journal} {\prl}\ }%
  \textbf{\bibinfo {volume} {100}},\ 
\bibinfo {pages} {235301} 
   (\bibinfo {year} {2008}).



   
\bibitem{spaceShuttle}M.\ Barmatz,  I.\ Hahn, J.~A.\ Lipa, and R.~V.\ Duncan,
\rmp  {\bf 79},1 (2007).



   
   \bibitem{West}%
  \BibitemOpen
  \bibfield{author}{%
  \bibinfo {author} {\bibfnamefont{J.~T.}~\bibnamefont{West}}, 
  \bibinfo {author} {\bibfnamefont{O.}\ \bibnamefont{Syshchenko}}, 
  \bibinfo {author} {\bibfnamefont{J.}\ \bibnamefont{Beamish}},\ and\ 
  \bibinfo {author} {\bibfnamefont{M.~H.~W.}\ \bibnamefont{Chan}},\ }%
  \bibfield{journal}{%
  {\bibinfo {journal} {Nature Physics}}\ }%
  \textbf{\bibinfo {volume} {5}},\ 
\bibinfo {pages} {598} 
(\bibinfo {year} {2009}).



\bibitem{ClarkWestChan2007}A.~C.\ Clark, J.~T.\ West, and M.~H.~W.\ Chan, 
\prl  {\bf 99}, 135302 (2007).

   
   

\bibitem{Lin2007}%
  \BibitemOpen
  \bibfield{author}{%
  \bibinfo {author} {\bibfnamefont{X.}~\bibnamefont{Lin}}, 
  \bibinfo {author}  {\bibfnamefont{M.~H.~W.}\ \bibnamefont{Chan}},\ and\ 
  \bibinfo {author} {\bibfnamefont{A.~C.}\ \bibnamefont{Clark}},\ }%
  \bibfield{journal}{%
  {\bibinfo {journal} {Nature}}\ }%
  \textbf{\bibinfo {volume} {449}},\ 
\bibinfo {pages} {1025} 
(\bibinfo {year} {2007}).


\bibitem{Lin2009}%
  \BibitemOpen
  \bibfield{author}{%
  \bibinfo {author} {\bibfnamefont{X.}~\bibnamefont{Lin}}, \bibinfo {author}
  {\bibfnamefont{A.~C.}\ \bibnamefont{Clark}}, \bibinfo {author}
  {\bibfnamefont{Z.~G.}\ \bibnamefont{Cheng}},\ and\ \bibinfo {author}
  {\bibfnamefont{M.~H.~W.}\ \bibnamefont{Chan}},\ }%
  \bibfield{journal}{%
  {\bibinfo {journal} {\prl}}\ }%
  \textbf{\bibinfo {volume} {102}},\ 
\bibinfo {pages} {125302} 
(\bibinfo {year} {2009}).

\bibitem{Hasenbusch}M.\ Campostrini, M.\ Hasenbusch, A.\ Pelissetto, and E.\ Vicari, 
\prb  {\bf 74}, 144506 (2006).

\bibitem{Harris}A.\ Harris, J.\ Phys.\ C: Solid State Phys.\ {\bf 7}, 1671 (1974).


\bibitem{Toner2008}%
  \BibitemOpen
  \bibfield{author}{%
  \bibinfo {author} {\bibfnamefont{J.}~\bibnamefont{Toner}},\ }%
  \bibfield{journal}{%
  \bibinfo {journal} {\prl}\ }%
  \textbf{\bibinfo {volume} {100}},\ 
\bibinfo {pages} {035302} 
   (\bibinfo {year} {2008}).



\bibitem{Goswami2011}%
  \BibitemOpen
  \bibfield{author}{%
  \bibinfo {author} {\bibfnamefont{D.}~\bibnamefont{Goswami}}, \bibinfo
  {author} {\bibfnamefont{K.}~\bibnamefont{Dasbiswas}}, \bibinfo {author}
  {\bibfnamefont{C.-D.}\ \bibnamefont{Yoo}},\ and\ \bibinfo {author}
  {\bibfnamefont{A.~T.}~\bibnamefont{Dorsey}},\ }%
  \bibfield{journal}{%
\bibinfo {journal} {\prb}\ }%
  \textbf{\bibinfo {volume} {84}},\ 
\bibinfo {pages} {054523} 
(\bibinfo {year} {2011}).


\bibitem{ProkofevRMP2012}M.\ Boninsegni and N.~V.\ Prokof'ev, 
\rmp {\bf 84}, 759 (2012).



\bibitem{Weinrib}A.\ Weinrib and B.~I.\ Halperin, \prb {\bf 27}, 413 (1983).



\bibitem{Li:1990p4584}%
  \BibitemOpen
  \bibfield{author}{%
  \bibinfo {author} {\bibfnamefont{Y.~H.}~\bibnamefont{Li}}\ and\ \bibinfo
  {author} {\bibfnamefont{S.}~\bibnamefont{Teitel}},\ }%
  \bibfield{journal}{%
  \bibinfo {journal} {\prb}\ }%
  \textbf{\bibinfo {volume} {41}},\ 
\bibinfo {pages} {11388} 
(\bibinfo {year}  {1990}).



\bibitem{FSS}V.~Privman, in {\it Finite Scaling and Numerical Simulations of Statistical Systems}, 
edited by V.~Privman (World Scientific, Singapore, 1990). 


\bibitem{wolff}%
  \BibitemOpen
  \bibfield{author}{%
  \bibinfo {author} {\bibfnamefont{U.}~\bibnamefont{Wolff}},\ }%
  \bibfield{journal}{%
  \bibinfo {journal} {\prl}\ }%
  \textbf{\bibinfo {volume} {62}},\ 
\bibinfo {pages} {361} 
(\bibinfo {year}  {1989}).



\bibitem{NumericalRecipesChpt15.6}W.~H.\ Press, S.~A.\ Teukolsky, W.~T.\ Vetterling and B.~P.\ Flannery, {\it Numerical Recipes}  3rd ed.\ (Cambridge University Press, 2007).


\bibitem{Ma}S.-K.\ Ma, {\it Modern Theory of Critical Phenomena} (Frontiers in Physics, 1976).


\end{thebibliography}
\end{document}